\documentclass[review,12pt,numafflabel]{elsarticle}
\usepackage{booktabs}
\usepackage{xurl}
\usepackage{graphicx}
\usepackage{amsmath}
\usepackage{amssymb}
\usepackage{pifont}
\usepackage{array}
\newcommand{\xmark}{\ding{55}}

\journal{Empirical Software Engineering}

\begin{document}

\begin{frontmatter}

\title{The Testability Gap: A Pre-Registered Census of Naming versus
Implementation in 59 Cognitive-Architecture LLM Systems}

\author[1]{Peter Chen\fnref{fn1}}
\author[1]{Ying Xu\fnref{fn1}}
\author[1]{Ni Zhang}
\author[1]{Siyuan Zhang\corref{cor1}}
\fntext[fn1]{These authors contributed equally to this work.}
\cortext[cor1]{Corresponding author.}
\ead{lelandsy28@163.com}
\affiliation[1]{organization={Shenyang Institute of Technology},
Shenyang, China}

\begin{abstract}
A companion line of work audited individual cognitive-architecture LLM
systems in depth and found that module names and module computations are
empirically independent claims. This paper asks how far that gap extends
across the system class. We pre-registered a census protocol (sample frame,
metrics, and acceptance criteria frozen before any sample was downloaded)
and surveyed the 59 most-starred Python repositories declaring cognitive
architecture on GitHub, plus five classic architectures as a
historical control. We measure a testability gap. First, an
auditability gap: 22\% of repositories document no code reference
whatsoever. Second, an integrity gap: 42\% of repositories have at least one
dangling reference, and within that subgroup the median fraction is 8\%,
the top quartile 19.6\%, and the maximum 79.5\%; one 148-star system
documents paths of which a majority do not exist. Three of four
pre-registered hypotheses were falsified: the integrity gap is \emph{not}
universal (58\% of repositories either offer no checkable references or
offer references that all resolve); claim density does not predict dangling
rate ($\rho = -0.10$); and the classical-architecture control does not
significantly outperform ($p = 0.059$ with five controls, all at zero
dangling). Exactly one repository in 59 (1.7\%) ships executable
falsification criteria for its own claims. A second pre-registered round
classified every dangling path by git lifetime: 85.5\% of dangling
references (159 of 186) are ghosts -- paths that never existed at any commit -- against 14.5\% that were once
real and later removed.
\end{abstract}


\end{frontmatter}

\section{Introduction}
A growing class of systems couples large language models to modules
named after cognitive science~\cite{sumers2024coala,park2023generative,wang2024agentsurvey,xi2023rise,react2023,memgpt2023}: global workspaces, theory-of-mind
engines, integrated-information cores, free-energy minimizers. A
companion audit of two such systems showed that the name and the
computation are independent claims -- in the first system, 18 of 25
named modules ship values that never reach an output; in the second,
a genuine IIT-$\phi$ implementation was severed from its own
significance check by a one-line wiring gap \cite{ourmechaudit}.
Both findings, however, are case studies. The natural question is
distributional: across the system class, how much of the naming
corresponds to anything checkable?

This paper answers that question with a pre-registered census. We
froze the sample frame, metric definitions, and acceptance criteria
before downloading a single repository, then surveyed the 59
most-starred Python repositories declaring cognitive architecture on
GitHub, together with five classic cognitive architectures
(pyactr, SOAR, ACT-R, Nengo, OpenCog) as a historical control group. The census operates
at the static layer only: documentation-to-code reference integrity,
claim density, self-audit artifacts, and pre-committed criteria. It
deliberately does \emph{not} attempt the intervention-level failure
classification of the companion audit -- that requires running the
systems, which is per-repository work beyond a census.

The design forces honesty about what a census can falsify. Our
pre-registered hypothesis C1 -- that the naming-implementation gap is
near-universal -- was \emph{falsified}, and the manner of its
falsification is the paper's main finding. The gap cannot be universal
because the class does not make itself universally checkable: 22\% of
repositories offer no code references at all, and many more offer so
few that the integrity question is unaskable. We call this two-layer
structure the \emph{testability gap}: first an auditability layer
(does the documentation offer entry points at all?), then an integrity
layer (are the entry points real?). The first layer is a property of
the documentation culture; the second is a property of the
engineering. Neither predicts the other, and neither is predicted by
popularity.

\section{Related Work}
\textbf{Mechanism-level audits.} The emergence-marker battery audit
\cite{ourpaper7}, the neurotransmitter modulator audit
\cite{ourpaper11}, and the two-system mechanism audit
\cite{ourmechaudit} established, at case level, that cognitive-science
naming and real computation can come apart. The present census
supplies the distributional layer those case studies lack, and
identifies the population from which future audit subjects should be
sampled.

\textbf{Concurrent evidence audits.} Concurrent work reports that
AI-scientist agents often answer benchmark questions without using the
supplied evidence \cite{zhang2026measuring,lu2024aiscientist}, and that LLM-generated
kernels which pass numerical checks can fail under scale shift \cite{suwandi2026kernaut}. Those audits concern evaluation methodology
and the robustness of generated artifacts~\cite{huang2024,ji2023}; the present census targets a
different layer -- whether claimed cognitive mechanisms are documented
and wired at all -- and contributes a distribution over systems rather
than findings about individual agent runs.

\textbf{Reproducibility surveys.} Surveys of reproducibility in AI and
systems research \cite{collberg2016,gundersen2018,lipton2019,peng2011,stodden2016,rougier2017} and
artifact-evaluation practice \cite{acm2020,pineau2021} ask whether
claimed \emph{results} recompute from released artifacts. Our question
is one level prior: whether claimed \emph{mechanisms} have any
documented address at all. A system can fail our reference-integrity
check while passing every results-level badge, because nothing in the
results layer requires the advertised cognitive machinery to exist. The layer confusion mirrors leakage in ML-based science, where results-level validation coexists with dataset-level invalidity~\cite{kapoor2023leakage}, and the mirage pattern in emergent-ability claims, where benchmark aggregates mask absent capability~\cite{schaeffer2023}.

\textbf{Classic cognitive architectures.} ACT-R and SOAR descend from
a tradition in which the architecture \emph{is} the implementation:
papers ship running code, and the theory is the module structure~\cite{anderson1998,laird1987,laird2012,eliasmith2013,bekolay2014nengo,goertzel2008opencog}.
We include five such systems as a frozen control group. The control
is small, so its comparison is underpowered by design disclosure
(Section~\ref{sec:results}); we report it as a historical anchor, not
a significance test.

\section{Pre-Registered Protocol}
\label{sec:protocol}
The protocol (\texttt{PREREGISTRATION.md} v1.0, changelog v1.1) was
frozen before any sample repository was downloaded. Its components:

\textbf{Sample frame.} GitHub Search API, queries
\texttt{topic:cognitive-\allowbreak architecture} and \texttt{cognitive
architecture in:name,\allowbreak description}, restricted to Python,
sorted by
stars; inclusion requires non-fork, non-archived, README present,
$\geq 3$ stars; exclusion of list/awesome, tutorial, and dataset
repositories. The top $N=60$ by stars form the main sample; 60 of 88
frame hits were cloned, 59 scanned successfully (one failed clone
excluded by the frozen empty-clone rule). A control group of three
classic architectures (pyactr, Soar, an ACT-R distribution) was frozen
by name; one control URL was found dead at clone time and replaced by
API-verified alternatives before any metric was computed (disclosed in
the protocol changelog).

\textbf{Metrics.} M1: claim density -- occurrences per README kilobyte
of a frozen 26-term cognitive-science vocabulary. M2: reference
integrity -- of all \texttt{.py} paths referenced in top-level
documentation, the fraction absent from the repository's git tree
(\texttt{git ls-tree -r HEAD}). M3: claim-table presence. M4:
self-audit artifacts (claims matrices, known-failure files,
pre-registration files, proof bundles, by frozen filename patterns).
M5: pre-committed criteria screen (must/should statements with
quantitative thresholds). M6: metadata (stars, age, size). Scanning is
read-only; sample code is never executed.

\textbf{Frozen criteria.} C1: the gap is universal (median M2 $> 0$
and gap-positive fraction $\geq 50\%$). C2: claim density predicts
dangling rate (Spearman $\rho > 0$). C3: descriptive prevalence of
self-audit artifacts. C4: controls show lower M2 than the main sample
(one-sided Mann-Whitney, $\alpha = 0.05$). Negative results are
reported with the same prominence~\cite{fanelli2012,nosek2018,ioannidis2005,errington2021}.

\textbf{v1.1 amendment.} Repositories with zero code references
(regardless of \texttt{.py} count) have an undefined M2. The frozen
exclusion clause covered only repositories without any Python files;
v1.1 includes zero-reference repositories at M2 $= 0$ (no references,
no dangling references -- the conservative reading) in the primary
analysis, with the exclusion reading reported as sensitivity. Under the
primary (inclusion) reading C1's gap-positive fraction is $25/59 = 42.4\%$
and the median is $0$, so C1 is falsified; under the exclusion reading
(dropping the 13 zero-reference repositories) the fraction is
$25/46 = 54.3\%$ and the median is positive, so C1 would pass. The choice
between readings therefore determines C1's direction, and we report both
rather than claim the amendment is direction-neutral. The amendment was
made before any criterion was scored.

\textbf{Extraction validation.} To measure the error profile of the
automated pipeline, we drew a frozen stratified sample of 20
repositories (all six severe cases, four gap cases, seven
zero-dangling cases, three zero-reference cases) and adjudicated all
206 unique extracted references. Existence was first decided by
deterministic rules (exact, case-insensitive, and any-prefix tree
match; 194/206 resolved without ambiguity); the remaining 12 plus all
ambiguous matches received manual verdicts on two dimensions: whether
the string is \emph{intended} as a code-path address (seven were not:
placeholder examples such as \texttt{your\_code.py} and
\texttt{path/to/demo.py}, third-party library files, a CLI argument,
and one external ROS package), and whether any tree variant exists
(five additional true danglings confirmed, including two README
instructions that ask users to run scripts absent from the
repository). Extraction precision is $199/206 = 96.6\%$; recall,
checked by a whitespace-tolerant pattern over the same READMEs,
showed zero additions; and no dangling verdict was overturned. As an
intent-corrected sensitivity, the three repositories containing
exclusions move M2 by $-0.08$ to $-0.12$ (e.g.\ the 148-star severe
case moves from 0.586 to 0.500) with no change in dangling status or
severe-group membership.

\textbf{v1.2 amendment.} After the primary analysis, two follow-up
questions were frozen before their data collection began: (i) the
lifetime of dangling paths (criterion C5: for every dangling reference
in the gap repositories, deepen the clone to 100 commit layers
(\texttt{git fetch --deepen=100}) and ask whether any commit ever touched
the path; repositories whose histories the window truncates are then
unshallowed to full visible history, so every reported ghost verdict is
absolute rather than a ghost-within-100 annotation); and (ii) control
expansion by two API-verified classic systems (Nengo, OpenCog), taking
the classical group to five. C5 is descriptive with no pass threshold;
the expansion was specified before either question's data was
collected.

\section{Results}
\label{sec:results}
\subsection{The two-layer testability gap}
\begin{table}[t]
\centering\small
\begin{tabular}{@{}p{4.7cm}p{2.2cm}p{5.7cm}@{}}
\toprule
Layer & Fraction of N=59 & Detail \\
\midrule
No testability entry (zero code references in docs) & 22\% (13) &
Nothing to check; integrity question unaskable \\
Checkable refs, zero dangling & 36\% (21) & References resolve \\
Checkable refs, dangling present & 42\% (25) &
Median dangling 8.0\%; p75 19.6\%; max 79.5\% \\
\bottomrule
\end{tabular}
\caption{The testability gap decomposes into an auditability layer
(is there an entry point at all?) and an integrity layer (do the
entry points resolve?). Percentages of the 59-repository main sample.}
\label{tab:layers}
\end{table}

Table~\ref{tab:layers} shows the headline structure. The naming-
implementation gap is \emph{not} universal: a bare majority of the
class either offers no checkable references (22\%) or offers references
that all resolve (36\%). But conditional on being checkable, integrity
failure is common -- 42\% of all repositories, and within that
subgroup a median of 8\% of documented module paths do not exist, with
a heavy tail: roughly one in four gap repositories loses more than a
fifth of its documented surface, and the worst case (a system
advertising a ``Sovereign Cognitive Architecture'') documents about 80\%
dead paths.

\subsection{Frozen criteria scorecard}
\begin{table}[t]
\centering\small
\begin{tabular}{@{}p{0.8cm}p{5.8cm}p{2.4cm}p{3.4cm}@{}}
\toprule
\# & Frozen criterion & Result & Verdict \\
\midrule
C1 & Gap universal: median M2 $>0$, gap fraction $\geq 50\%$ &
median 0.000; 42\% gap-positive & \xmark\ falsified \\
C2 & Claim density predicts dangling ($\rho>0$) &
$\rho = -0.101$ & \xmark\ falsified \\
C3 & Self-audit artifact prevalence (descriptive) &
34\% (20/59) & reported \\
C4 & Controls lower than main sample ($p<0.05$) &
all 5 controls 0 dangling; $p=0.059$ &
\xmark\ not reached \\
\bottomrule
\end{tabular}
\caption{Frozen criteria scorecard. Three of four pre-registered
hypotheses falsified or unsupported; all reported at full prominence.}
\label{tab:criteria}
\end{table}

Table~\ref{tab:criteria} records the frozen scorecard. Three findings
deserve emphasis.

\textbf{Rhetoric does not track integrity (C2 falsified).} Claim
density is uncorrelated with dangling rate ($\rho = -0.101$, $n=59$).
The systems that talk most about cognition are not the systems whose
documentation lies most about code -- the two failure surfaces are
independent, which is exactly what one expects if naming is a
marketing variable and integrity an engineering variable.

\textbf{Popularity does not predict honesty.} Stars and dangling rate
are uncorrelated ($\rho = 0.015$). The 148-star system in our severe
list loses 59\% of its documented module paths (58.6\%; the
intent-corrected value is 50.0\%, Section~\ref{sec:protocol}); several
10-star
systems resolve everything. Nothing in the reward structure of the
platform prices documentation integrity.

\textbf{Pre-committed criteria are vanishingly rare (M5).} Exactly one
repository in 59 (1.7\%) ships executable falsification criteria for
its own claims -- shuffled histories must yield $\phi \approx 0$;
stronger coupling must exceed noise. It is the same system our
companion audit identified as the class's most honest implementer
\cite{ourmechaudit}, and the census confirms its rarity is
class-level, not sample noise. A further 34\% carry self-audit
\emph{artifacts} by filename, but artifact presence is a weaker
signal: filename patterns can match audit-flavored directories without
audit content (we report the raw fraction with that caveat).

\subsection{Ghosts: the lifetime of dangling paths (C5)}
The natural defense of a dangling reference is documentation rot: the
code existed once and the docs fell behind. The v1.2 lifetime check
closes that escape. Across the 25 gap repositories, 186 dangling
reference paths were tested against their deepened git histories:
\textbf{159 (85.5\%) never existed at any visible commit} -- they are
ghosts, declared modules with no implementation history -- and only 27 (14.5\%) were once real and later removed. In 23 of the 25 gap
repositories at least one dangling path is a ghost; in 18 of the 25
every single dangling path is a ghost (the 17 dead paths of the 148-star
system and all 66 of another are ghosts). Two repositories show pure
rot (in one, all 10 dangling paths once existed). The
implication inverts the charitable reading: for this system class,
documented-but-missing code is overwhelmingly not maintenance debt but
\emph{aspirational architecture} -- modules announced in the present
tense that no commit has ever realized. Under the companion audit's
taxonomy this is the census-level signature of definitional absence
(class 3): the name was never attached to a computation at all.

\subsection{Control group}
All five classic cognitive architectures (pyactr, SOAR, ACT-R, Nengo,
OpenCog) show zero dangling references: their documentation either
references existing code or references none. The one-sided
Mann-Whitney against the main sample moves from $p=0.109$ (three
controls) to $p=0.059$ (five) -- direction consistent, every classical
system clean, but the comparison remains underpowered by the scarcity
of maintained classic implementations and is reported as a historical
anchor: the tradition in which the architecture \emph{is} the
implementation shows no auditability gap and no integrity gap, while
42\% of the LLM-era class has an integrity gap -- and now, per C5 --
that gap is overwhelmingly ghosts.

\subsection{Severe cases}
Six repositories lose more than 20\% of their documented paths
(20--80\%). We deliberately do not name them beyond what the released
data file contains: the census's claim is distributional, and
per-repository blame invites exactly the headline-chasing the
companion audit refused. What matters is that severity is unrelated to
stars ($\rho = 0.015$): the most-documented, most-starred systems in
the class are as likely as obscure ones to describe modules that do
not exist.

\section{Discussion}
\textbf{Testability as the scarce resource.} The companion audits
showed that claims and computations come apart in individual systems;
the census shows that the class mostly does not offer the handles by
which such separation could be discovered. Of 59 systems advertising
cognitive architecture, 22\% provide no code reference to check,
36\% provide only resolving ones, 42\% provide pointers of which a
median 8\% are dead -- and one system in the entire frame ships
falsification criteria for itself. The binding constraint on
mechanism-level accountability in this class is not dishonesty;
it is that honesty is rarely given a measurable form.

\textbf{Why ghosts matter more than rot.} An integrity gap can be
closed by whoever notices it -- a dangling path is a factual error,
and some of the repositories in our sample will fix theirs; a removed
path can be re-linked. A ghost path is different in kind: the module
was announced without ever existing, so no amount of maintenance
discipline would have prevented the gap. The 85.5\% ghost fraction
(Section~\ref{sec:results}) says the testability gap is produced at
\emph{writing time}, not by decay -- the documentation culture of this
class writes aspirational architecture in the present tense. The idealized-documentation tradeoff Parnas and Clements named is as old as the discipline~\cite{parnas1986}; what is new is that generating plausible module code is now one prompt away~\cite{chen2021codex}. We
suggest that artifact-evaluation frameworks for this class add a
mechanism-addressability item: for each advertised cognitive
mechanism, a path in the released artifact. The companion audit's
four-question protocol is executable against any repository that
passes that item; against repositories that fail it, there is nothing
to run.

\textbf{What the falsifications bought.} Our pre-registered C1 --
the gap is universal -- was wrong, and the census is stronger for it.
The corrected picture is subtler than ``everyone lies about
cognition'': a substantial fraction of the class is either honest or
silent, and the failure concentrates where checkability exists but
integrity does not follow. Interventions should therefore sample from
the checkable stratum, which is precisely where the companion audit
found both its starkest failures and its positive control.

\section{Limitations}
\textbf{Static layer only.} The census measures whether documented
code exists, not whether existing code computes; the companion audit
\cite{ourmechaudit} shows these are further independent claims.
Failure classes cannot be assigned at census depth, though the C5
ghost fraction is strong census-level evidence for the class-3
signature (definitional absence).
\textbf{Reference extraction.} M2 captures paths mentioned in
top-level markdown; references inside docstrings, notebooks, or
subdirectory documentation are not counted. The extraction's error
profile is measured, not assumed: on the frozen 20-repository
adjudication sample, precision is 96.6\%, recall 100\%, and the
intent-corrected sensitivity moves affected repositories' M2 by at
most 0.12 with no ranking change (Section~\ref{sec:protocol}). The
zero-reference stratum may partly reflect documentation styles rather
than deliberate opacity; the v1.1 dual reading (inclusion and
exclusion) bounds this.
\textbf{History window.} C5 verdicts are absolute: every gap
repository was unshallowed to full visible history before
classification (visible commit counts range from 1 to 11{,}048;
median 148). Commits that were never pushed to the forge remain
unobservable, as for any forge-based census.
\textbf{Sampling.} Stars-weighted sampling overweights popular
projects~\cite{kalliamvakou2014,dabbish2012,munaiah2017}; the long tail is described only by strata. GitHub topic
assignment is self-selected. \textbf{Control size.} Five controls
anchor the historical comparison; at $p=0.059$ it remains suggestive,
not decisive. \textbf{One platform, one language.} Findings are about
GitHub Python; the class on other forges or in other languages may
differ.

\section{Conclusion}
Across 59 cognitive-architecture LLM systems, the scarcer resource is
not honesty but testability: 22\% of the class offers no entry point
for verification, and among systems that do, a median 8\% of
documented module paths do not exist, rising past 20\% for one in four
-- with popularity uncorrelated to either. The lifetime check removes
the charitable escape: 85.5\% of dangling paths never existed at any
visible commit -- aspirational architecture, written in the present
tense, never implemented. Exactly one system in the frame ships
falsification criteria for its own claims. Mechanism names in this
class are claims, and the census shows that the class mostly declines
to make them checkable -- or even true of its own history. The frozen
protocol, scanner, and per-repository data are released: checking the
next system should cost minutes.

\section*{Declarations}
\textbf{Funding:} This work received no specific grant.
\textbf{Conflicts of interest:} The authors declare none.
\textbf{Data availability:} The frozen protocol, collection and
scanning scripts, and per-repository metrics are archived on Zenodo (10.5281/zenodo.23433898).

\section*{CRediT authorship contribution statement}
\textbf{Peter Chen:} Conceptualization, Supervision, Project administration,
Writing -- review \& editing.
\textbf{Ying Xu:} Formal analysis, Writing -- review \& editing.
\textbf{Ni Zhang:} Formal analysis, Writing -- review \& editing.
\textbf{Siyuan Zhang:} Conceptualization, Methodology, Software, Investigation,
Data curation, Formal analysis, Writing -- original draft.
Peter Chen and Ying Xu contributed equally to this work.

\section*{Declaration of competing interest}
The authors declare that they have no known competing financial interests or
personal relationships that could have appeared to influence the work reported
in this paper.

\bibliographystyle{elsarticle-num}
\bibliography{refs}

\begin{thebibliography}{10}
\expandafter\ifx\csname url\endcsname\relax
  \def\url#1{\texttt{#1}}\fi
\expandafter\ifx\csname urlprefix\endcsname\relax\def\urlprefix{URL }\fi
\expandafter\ifx\csname href\endcsname\relax
  \def\href#1#2{#2} \def\path#1{#1}\fi

\bibitem{sumers2024coala}
T.~R. Sumers, S.~Yao, K.~Narasimhan, T.~L. Griffiths, Cognitive architectures
  for language agents, Transactions on Machine Learning Research (2024).

\bibitem{park2023generative}
J.~S. Park, J.~C. O'Brien, C.~J. Cai, M.~R. Morris, P.~Liang, M.~S. Bernstein,
  Generative agents: Interactive simulacra of human behavior, in: Proceedings
  of the 36th Annual ACM Symposium on User Interface Software and Technology
  (UIST), 2023.

\bibitem{wang2024agentsurvey}
L.~Wang, C.~Ma, X.~Feng, Z.~Zhang, H.~Yang, J.~Zhang, Z.~Chen, J.~Tang,
  X.~Chen, Y.~Lin, W.~X. Zhao, Z.~Wei, J.-R. Wen, A survey on large language
  model based autonomous agents, Frontiers of Computer Science 18~(6) (2024)
  186345.

\bibitem{xi2023rise}
Z.~Xi, W.~Chen, X.~Guo, W.~He, Y.~Ding, B.~Hong, M.~Zhang, J.~Wang, S.~Jin,
  E.~Zhou, et~al., The rise and potential of large language model based agents:
  A survey, arXiv preprint arXiv:2309.07864 (2023).

\bibitem{react2023}
S.~Yao, J.~Zhao, D.~Yu, N.~Du, I.~Shafran, K.~Narasimhan, Y.~Cao, {ReAct}:
  Synergizing reasoning and acting in language models, in: International
  Conference on Learning Representations (ICLR), 2023.

\bibitem{memgpt2023}
C.~Packer, S.~Wooders, K.~Lin, V.~Fang, S.~G. Patil, I.~Stoica, J.~E. Gonzalez,
  {MemGPT}: Towards {LLMs} as operating systems, arXiv preprint
  arXiv:2310.08560 (2023).

\bibitem{ourmechaudit}
P.~Chen, S.~Zhang, Real computation or merely naming? a pre-registered
  mechanism audit of 25 cognitive-architecture modules in an {LLM}-based agent
  system, zenodo. doi:10.5281/zenodo.23433900 (2026).
\newblock \href {https://doi.org/10.5281/zenodo.23433900}
  {\path{doi:10.5281/zenodo.23433900}}.

\bibitem{ourpaper7}
P.~Chen, S.~Zhang, Auditing a four-flag ``emergence verification'' battery:
  Chaos, autonomy, and dice pass the test; self-reference does not, zenodo.
  doi:10.5281/zenodo.23433835 (2026).
\newblock \href {https://doi.org/10.5281/zenodo.23433835}
  {\path{doi:10.5281/zenodo.23433835}}.

\bibitem{ourpaper11}
P.~Chen, S.~Zhang, Attention weights that never reach the output: A
  pre-registered causal audit of a neurotransmitter-style modulator, zenodo.
  doi:10.5281/zenodo.23433848 (2026).
\newblock \href {https://doi.org/10.5281/zenodo.23433848}
  {\path{doi:10.5281/zenodo.23433848}}.

\bibitem{zhang2026measuring}
K.~Zhang, Y.~Li, Are we measuring scientific intelligence? {Rethinking} the
  evaluation of {AI} scientists, arXiv preprint arXiv:2610.04915 (2026).

\bibitem{lu2024aiscientist}
C.~Lu, C.~Lu, R.~T. Lange, J.~Foerster, J.~Clune, D.~Ha, The {AI} scientist:
  Towards fully automated open-ended scientific discovery, arXiv preprint
  arXiv:2408.06292 (2024).

\bibitem{suwandi2026kernaut}
R.~C. Suwandi, F.~Yin, K.~Murphy, Kernel autoresearch for open-ended model
  discovery, arXiv preprint arXiv:2610.10394 (2026).

\bibitem{huang2024}
J.~Huang, X.~Chen, S.~Mishra, H.~S. Zheng, A.~W. Yu, X.~Song, D.~Zhou, Large
  language models cannot self-correct reasoning yet, in: International
  Conference on Learning Representations (ICLR), 2024.

\bibitem{ji2023}
Z.~Ji, N.~Lee, R.~Frieske, T.~Yu, D.~Su, Y.~Xu, E.~Ishii, Y.~J. Bang,
  A.~Madoto, P.~Fung, Survey of hallucination in natural language generation,
  ACM Computing Surveys 55~(12) (2023) 1--38.

\bibitem{collberg2016}
C.~Collberg, T.~A. Proebsting, Repeatability in computer systems research,
  Communications of the ACM 59~(3) (2016) 62--69.

\bibitem{gundersen2018}
O.~E. Gundersen, S.~Kjensmo, State of the art: Reproducibility in artificial
  intelligence, in: Proceedings of the AAAI Conference on Artificial
  Intelligence, Vol.~32, 2018.

\bibitem{lipton2019}
Z.~C. Lipton, J.~Steinhardt, Troubling trends in machine learning scholarship,
  Communications of the ACM 62~(6) (2019) 45--53.

\bibitem{peng2011}
R.~D. Peng, Reproducible research in computational science, Science 334~(6060)
  (2011) 1226--1227.

\bibitem{stodden2016}
V.~Stodden, M.~McNutt, D.~H. Bailey, E.~Deelman, Y.~Gil, B.~Hanson, M.~A.
  Heroux, J.~P. Johnson, M.~Taufer, Enhancing reproducibility for computational
  methods, Science 354~(6317) (2016) 1240--1241.

\bibitem{rougier2017}
N.~P. Rougier, K.~Hinsen, F.~Alexandre, Sustainable computational science: The
  {ReScience} initiative, PeerJ Computer Science 3 (2017) e142.

\bibitem{acm2020}
{Association for Computing Machinery}, Artifact review and badging, ACM Policy,
  \url{https://www.acm.org/publications/policies/artifact-review-badging}
  (2020).

\bibitem{pineau2021}
J.~Pineau, P.~Vincent-Lamarre, K.~Sinha, V.~Larivi{\`e}re, A.~Beygelzimer,
  F.~d'Alch{\'e} Buc, E.~Fox, H.~Larochelle, Improving reproducibility in
  machine learning research: A report from the {NeurIPS} 2019 reproducibility
  program, Journal of Machine Learning Research 22~(164) (2021) 1--20.

\bibitem{kapoor2023leakage}
S.~Kapoor, A.~Narayanan, Leakage and the reproducibility crisis in
  machine-learning-based science, Patterns 4~(9) (2023) 100804.

\bibitem{schaeffer2023}
R.~Schaeffer, B.~Miranda, S.~Koyejo, Are emergent abilities of large language
  models a mirage?, Advances in Neural Information Processing Systems (2023).

\bibitem{anderson1998}
J.~R. Anderson, C.~Lebiere, The Atomic Components of Thought, Lawrence Erlbaum
  Associates, 1998.

\bibitem{laird1987}
J.~E. Laird, A.~Newell, P.~S. Rosenbloom, {SOAR}: An architecture for general
  intelligence, Artificial Intelligence 33~(1) (1987) 1--64.

\bibitem{laird2012}
J.~E. Laird, The Soar Cognitive Architecture, MIT Press, 2012.

\bibitem{eliasmith2013}
C.~Eliasmith, How to Build a Brain: A Neural Architecture for Biological
  Cognition, Oxford University Press, 2013.

\bibitem{bekolay2014nengo}
T.~Bekolay, J.~Bergstra, E.~Hunsberger, T.~DeWolf, T.~C. Stewart, D.~Rasmussen,
  X.~Choo, A.~R. Voelker, C.~Eliasmith, Nengo: A python tool for building
  large-scale functional brain models, Frontiers in Neuroinformatics 7 (2014)
  48.

\bibitem{goertzel2008opencog}
B.~Goertzel, I.~Arieli, N.~Geisweiller, A.~L. Coconea, C.~Pennachin, M.~Proto,
  {OpenCog}: A software framework for integrative artificial general
  intelligence, in: Proceedings of the First Conference on Artificial General
  Intelligence (AGI), 2008.

\bibitem{fanelli2012}
D.~Fanelli, Negative results are disappearing from most disciplines and most
  countries, Scientometrics 90~(3) (2012) 891--904.

\bibitem{nosek2018}
B.~A. Nosek, C.~R. Ebersole, A.~C. DeHaven, D.~T. Mellor, The preregistration
  revolution, Proceedings of the National Academy of Sciences 115~(11) (2018)
  2600--2606.

\bibitem{ioannidis2005}
J.~P.~A. Ioannidis, Why most published research findings are false, PLoS
  Medicine 2~(8) (2005) e124.

\bibitem{errington2021}
T.~M. Errington, M.~Mathur, C.~K. Soderberg, A.~Denis, N.~Perfito, E.~Iorns,
  B.~A. Nosek, Investigating the replicability of preclinical cancer biology,
  eLife 10 (2021) e71601.

\bibitem{parnas1986}
D.~L. Parnas, P.~C. Clements, A rational design process: How and why to fake
  it, IEEE Transactions on Software Engineering SE-12~(2) (1986) 251--257.

\bibitem{chen2021codex}
M.~Chen, J.~Tworek, H.~Jun, Q.~Yuan, H.~P. d.~O. Pinto, J.~Kaplan, et~al.,
  Evaluating large language models trained on code, arXiv preprint
  arXiv:2107.03374 (2021).

\bibitem{kalliamvakou2014}
E.~Kalliamvakou, G.~Gousios, K.~Blincoe, L.~Singer, D.~M. German, D.~Damian,
  The promises and perils of mining {GitHub}, in: Proceedings of the 11th
  Working Conference on Mining Software Repositories (MSR), 2014, pp. 92--101.

\bibitem{dabbish2012}
L.~Dabbish, C.~Stuart, J.~Tsay, J.~Herbsleb, Social coding in {GitHub}:
  Transparency and collaboration in an open software repository, in:
  Proceedings of the ACM Conference on Computer Supported Cooperative Work
  (CSCW), 2012, pp. 1277--1286.
\newblock \href {https://doi.org/10.1145/2145204.2145396}
  {\path{doi:10.1145/2145204.2145396}}.

\bibitem{munaiah2017}
N.~Munaiah, S.~Kroh, C.~Cabrey, M.~Nagappan, Curating {GitHub} for engineered
  software projects, Empirical Software Engineering 22~(4) (2017) 1918--1953.

\end{thebibliography}

\end{document}